\documentclass[12pt]{article}
\usepackage{amssymb,amsmath,amsthm,graphicx,ulem}
\usepackage[linktoc=page]{hyperref}

\usepackage{graphicx,subfigure}
\usepackage{epsfig}
\usepackage{amsmath}
\usepackage{amsfonts}
\usepackage{amssymb}
\usepackage[usenames]{color}
\usepackage{pstricks,pst-coil,pst-plot,pst-3dplot,pst-node}
\usepackage{pst-grad}
\usepackage[letterpaper,left=2.cm,right=2.cm,top=2.5cm,bottom=2.5cm]{geometry}
\usepackage[T1]{fontenc}

\newcommand{\beq}{\begin{equation}}
\newcommand{\eeq}{\end{equation}}
\newcommand{\be}{\begin{equation}}
\newcommand{\ee}{\end{equation}}
\newcommand{\beqa}{\begin{eqnarray}}
\newcommand{\eeqa}{\end{eqnarray}}
\newcommand{\beqar}{\begin{eqnarray*}}
\newcommand{\eeqar}{\end{eqnarray*}}
\newcommand{\bea}{\begin{eqnarray}}
\newcommand{\eea}{\end{eqnarray}}

\numberwithin{equation}{section}

\numberwithin{equation}{section}

\begin{document}

\allowdisplaybreaks

\normalem

\title{Extremal Surface Barriers}

\author{Netta Engelhardt and Aron C. Wall
\\ 
\\ \\
  \textit{Department of Physics}\\
\textit{University of California, Santa Barbara}\\
\textit{ Santa Barbara, CA 93106, USA} \\ 
 \\ 
 \small{engeln@physics.ucsb.edu, aroncwall@gmail.com}}

 \date{}

 \maketitle

\begin{abstract}
\noindent We present a generic condition for Lorentzian manifolds to have a barrier that limits the reach of boundary-anchored extremal surfaces of arbitrary dimension.  
We show that any surface with nonpositive extrinsic curvature is a barrier, in the sense that extremal surfaces cannot be continuously deformed past it.
Furthermore, the outermost barrier surface has nonnegative extrinsic curvature.  
Under certain conditions, we show that the existence of trapped surfaces implies a barrier, and conversely.
In the context of AdS/CFT, these barriers imply that it is impossible to reconstruct the entire bulk using extremal surfaces.  We comment on the implications for the firewall controversy.
\end{abstract}

\newpage


\tableofcontents

\baselineskip16pt

\section{Introduction}\label{intro}
Despite the successes of the AdS/CFT correspondence, a full understanding of the duality is still lacking. Since the original conjecture \cite{MaldacenaAdSCFT}, much of the research effort has been aimed at developing a precise dual dictionary connecting geometric quantities in the bulk to observables in the boundary field theory. However, this dictionary remains poorly understood, hampering our ability to reconstruct the bulk geometry from the field theory, compute field theory quantities from the classical bulk theory, and assess whether or not the correspondence is complete. This last effort has been placed under scrutiny by the recent controversy over the black hole interior (see e.g.\cite{AMPS, AMPSS, MarolfPolchinski, PapadodimasRaju, Giddings, MaldacenaSusskind, Bousso, PapadodimasRaju2, AveryChowdhuryPuhm}), and direct efforts to construct -- or demonstrate an inability to construct -- a full dual dictionary, may serve to address it.

To resolve such questions, a good starting point is the identification of bulk probes that depend primarily on the bulk geometry and are dual by the AdS/CFT dictionary to known field theory observables. The existence of such bulk probes is not guaranteed; however, spacelike extremal surfaces, which are covariantly-defined and depend exclusively on the bulk geometry, are dual to well-understood field theory observables. In fact, spacelike extremal surfaces constitute most of the probes used in AdS/CFT. The dual observables include correlators \cite{KrausOoguriShenker}, entanglement entropy \cite{RyuTakayanagi, HubenyRangamaniTakayanagi}, and Wilson loops \cite{MaldacenaWLoop}. 
The length of spacelike geodesics with boundary endpoints allows one to compute, in the WKB approximation, the two-point correlator of a high conformal dimension field operator at the endpoints. The area of a codimension 2 spacelike extremal surface anchored on some region $\mathcal{R}$ on the boundary of an asymptotically locally Anti-de Sitter (AlAdS) spacetime is associated with the entanglement entropy of $\mathcal{R}$ within the boundary field theory. 

A field theory observable which is dual to an extremal surface must in some way encode information about the bulk geometry at the location of the extremal surface.  If there is a limit on how far into the bulk such surfaces can reach, this also limits our ability to reconstruct the geometry from the corresponding dual observables.  The natural question that arises is how much of the bulk can be recovered from extremal surface probes. Extremal surface probes and their maximal reach were studied in \cite{Hubeny} for the case of static and translationally or spherically symmetric asymptotically AdS spacetimes\footnote{A study on the reach of extremal surfaces in static black holes with higher derivative terms was done in \cite{Pal}.}, and in a large amount of literature (e.g. \cite{FidkowskiHubenyKlebanShenker} and \cite{HartmanMaldacena}) for eternal AdS black holes. Studies have also been done for extremal surfaces in time-dependent geometries (e.g. \cite{HubenyRangamaniTakayanagi, MaldacenaPimentel, EngelhardtHorowitz, LiuSuh, FischlerKundu, FischlerKunduPedraza}).

In this paper, we establish a general constraint on the reach of probe spacelike extremal surfaces in spacetimes which are not necessarily AlAdS.  We find that many generic geometries admit a certain kind of surface which acts as a barrier for probe surfaces anchored to one boundary. We furthermore show that trapped surfaces with nonpositive null extrinsic curvature give rise to extremal surface barriers, and that---on spacetimes which admit a totally geodesic spatial slice---the existence of barriers implies the existence of either a singularity or a marginally trapped surface.

More explicitly, we consider a codimension 1 ``splitting surface'' $\Sigma$ which divides the spacetime into two regions.  $\Sigma$ may be spacelike, timelike, or null.  We take some class $\cal X$ of spacelike extremal surfaces of codimension 1 or greater, which are anchored to some boundary $I$ on one side of $\Sigma$.  Then, given two simple assumptions, we can show that no members of this class of extremal surfaces cross (or even touch) $\Sigma$.  In other words, the splitting surface acts as an extremal surface barrier.  The two assumptions are as follows:
\begin{enumerate}
\item All of the extremal surfaces in the class $\cal X$ can be continuously deformed (while remaining extremal) to surfaces which lie on the exterior of $\Sigma$.
\item $\Sigma$ has nonpositive extrinsic curvature (as measured by normal vectors pointing outward, towards the exterior of $\Sigma$).  In other words, $\Sigma$ can only bend outward relative to its tangent plane.
\end{enumerate}
Given these assumptions, $\Sigma$ acts as a barrier to any of the extremal surfaces lying in the class $\cal X$.  This generalizes some previously known result in Riemannian manifolds, that surfaces with an inward-pointing mean curvature vector can sometimes act as local barriers to minimal surfaces, and also act as global barriers in certain circumstances (see e.g. \cite{MeeksYau, SchoenSimon, Spadaro}).

Our proof uses elements from the approach of \cite{Wall} and \cite{Wall2}, and shows that spacelike extremal surfaces cannot approach surfaces with negative extrinsic curvature. They can, however, approach arbitrarily close to totally geodesic surfaces. We do not need to make any assumptions about causality or an energy condition; the restriction is purely geometrical in nature.

The first criterion is often satisfied: it simply requires that there be no obstruction to deforming the extremal surface (while maintaining extremality) away from $\Sigma$ \footnote{For some field theory observables, the bulk extremal surface is required to be the \emph{minimal} area surface of those which are extremal.  This can cause the location of the surface to jump discontinuously as one changes where it is anchored to $I$ (e.g. geodesics in the BTZ black hole geometry \cite{RyuTakayanagi}).  This is not an issue for our proof, since we do not require that the deformed extremal surface be minimal. }.  In many cases, this criterion holds for all extremal surfaces due to the topology and geometry of the spacetime, so the class $\cal X$ consists of all extremal surfaces.  For example, this is true in vacuum AdS spacetimes.  More generally, if the extremal surfaces are anchored to just one connected component of the boundary $I$, and the bulk spacetime is homotopically trivial, then all surfaces may be continuously deformed towards $I$.  However, it may be an additional constraint on the geometry that the surfaces can remain extremal while this happens.  For example, if there is another region $\Sigma^\prime$ which acts as a barrier to extremal surfaces anchored anywhere on the boundary of the spacetime $\partial M$, then $\Sigma^\prime$ may be an obstruction to deforming extremal surfaces past $\Sigma$.






The second criterion may be satisfied by a totally geodesic surface, i.e. a surface with vanishing extrinsic curvature \footnote{A totally geodesic surface $\Sigma$ can also be equivalently defined as a surface $\Sigma$ in $(M,g)$ such that every geodesic on $\Sigma$ (with respect to the induced connection on $\Sigma$) is also a geodesic on $M$ (with respect to the connection on $M$. }.  An example of such a surface is a non--expanding black hole horizon (which, given the area theorem  \cite{Hawking}, is a stationary black horizon for spacetimes obeying the null energy condition). However, totally geodesic surfaces do not appear in generic spacetimes.  Fortunately, the result also applies to surfaces with negative extrinsic curvature, which can appear in generic spacetimes.  Examples will be given in Section \ref{examples}. In fact, if there exists a surface $\Sigma$ whose extrinsic curvature is negative everywhere, we will show that it is not actually the tightest bound on extremal surfaces.  There will be some other ``outermost'' barrier $\Sigma^\prime$ some distance away, which has at least partly nonnegative extrinsic curvature (in Section \ref{Singularities}, we will prove a theorem relating compact outermost barriers to trapped surfaces and singularities).



The existence of such extremal surface barriers and their link to singularities is a curiosity which is of relevance to the firewall controversy. The fact that 
many bulk probes do not extend beyond the barrier
may suggest that the firewall, if it exists, may be at or behind the 
outermost barrier.
At the very least, it shows that field theory observables that are dual to extremal surfaces anchored at one boundary do not directly reveal any information about the interior.  Any information about this region in the bulk must come from probes that are not extremal surfaces.  In some cases, the presence of a barrier might even indicate that the boundary field theory does not have 
sufficient information to reconstruct the geometry behind the barrier \cite{MarolfWall}.  In that case, additional factors of the Hilbert space may be needed to describe that region.

An interpretation involving a loss of determinism may at first seem implausible in the case where $\Sigma$ is spacelike: one could use the bulk equations of motion to evolve the information in $\Sigma$ forwards or backwards in time and thus reconstruct some or all of the interior of $\Sigma$. However, the presence of a firewall at $\Sigma$ would result in a breakdown of the classical bulk equations of motion. In this case, our hope of reconstructing the interior of $\Sigma$ relies entirely on the AdS/CFT correspondence dictionary. This dictionary, however, is primarily composed of extremal surface probes, which we prove below cannot probe past $\Sigma$.

This paper is structured as follows. In Section \ref{theorem}, we prove our sufficient condition for a spacetime to have a ``barrier surface'' past which extremal surfaces cannot reach; we further prove that trapped surfaces form extremal surface barriers. We will also show that the outermost barrier surface must have partly nonnegative extrinsic curvature.  Section \ref{examples} provides examples of bulk spacetimes with extremal surface barriers. In Section \ref{Singularities}, we prove that in a certain a class of geometries, barriers occur only in the presence of singularities or trapped surfaces.

\section{Barrier Theorems}\label{theorem}

We first prove in two parts that extremal surfaces do not probe past surfaces with a nonpositive extrinsic curvature tensor, and further, that barriers with everywhere negative extrinsic curvature are not the outermost barrier surfaces. $(M,g)$ will be a Lorentzian manifold with at least one boundary $I$ for the rest of this paper.  We start with two definitions. 

Define a codimension 1 surface $\Sigma$ in $M$ to be a \textit{splitting surface} if it separates $M$ into two open regions, Ext$(\Sigma)$ and Int$(\Sigma)$, where we define the interior and exterior relative to the normal which defines the extrinsic curvature, and $\partial \Sigma=\emptyset$. This normal is taken to point towards Ext$(\Sigma)$.

Let $\Sigma$ now be a splitting surface. Let $\{N_{r}\}$ be a family of spacelike extremal surfaces in $M$ of codimension $n\geq 1$ such that all the $N_{r}$ can all be continuously deformed from some initial surface $N_{0}\subset \text{Ext}(\Sigma)$ anchored at Ext$(\Sigma)\cap I$, and all the $N_{r}$ are anchored at Ext$(\Sigma) \cap I$ (see, e.g. Fig. \ref{def}).\footnote{For simplicity, we will assume that all extremal surfaces in the family are twice differentiable, at least where they intersect $\Sigma$, so that the extrinsic curvature $K_{\mu\nu}$ is well-defined.  Presumably this assumption can be weakened if care is taken in dealing with distributional extrinsic curvatures.} Then any surface in the family $N_{r}$ shall be called \textit{$\Sigma$-deformable}.

\subsection{$K < 0$ Surfaces are Barriers}

\textbf{Theorem 2.1:} Let $\Sigma$ be a codimension 1 splitting surface in $M$ such that for any vector field $v^{\mu}$ on $\Sigma$, $^{\Sigma}\!K_{\mu\nu}v^{\mu}v^{\nu}< 0$, where $^{\Sigma}\!K_{\mu\nu}$ is the extrinsic curvature of $\Sigma$ \footnote{In terms of  $\gamma_{\mu\nu}$, the induced metric on $\Sigma$, the extrinsic curvature on $\Sigma$ is given by $^{\Sigma}K_{\mu\nu} =\gamma^{\sigma}_{\mu}\gamma_{\lambda \nu}\nabla_{\sigma}\phantom{}^{\Sigma}k^{\lambda}$. For a surface $N$ of codimension $n>1$, the extrinsic curvature carries an additional index, and is defined in terms of $h_{\mu\nu}$, the induced metric on $N$: $^{N}K_{\mu\nu}^{\rho}=h_{\mu}^{\sigma}h_{\nu \lambda}\nabla_{\sigma}h^{\rho\lambda}$.}. Any $\Sigma$-deformable spacelike extremal surface which is anchored within $I \cap \mathrm{Ext}(\Sigma)$ remains in Ext$(\Sigma)$. Moreover, no such spacelike extremal surface ever touches $\Sigma$ (or even comes arbitrarily close to touching $\Sigma$). \footnote{If $^{\Sigma}\!K_{\mu\nu}v^{\mu}v^{\nu}>0$, we could prove, using precisely the same formalism used in the proof above, that surfaces anchored in Int$(\Sigma)$ remain in Int$(\Sigma)$.}

\begin{proof}
The proof closely follows that of Lemma B in \cite{Wall}. Let $N_{r}$ be the family of surfaces all deformable from $N_{0}$ as defined above. If all surfaces $N_{r}$ are in Ext$(\Sigma)$, we are done, since that implies that surfaces which are anchored at $I\cap \text{Ext}(\Sigma)$ can never be deformed past $\Sigma$. 

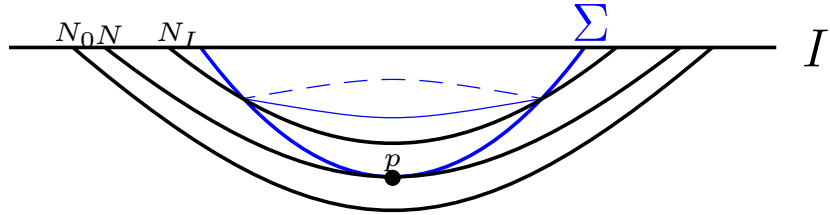
\begin{figure}
\begin{center}
\psscalebox{1.7}
{
\begin{pspicture}(-0.2,1)(5,4)

\psbezier[linecolor=blue](1,3)(2,1.65)(3,1.65)(4,3)
\psbezier[linecolor=black](0.25, 3)(2,1.65)(3,1.65)(4.75,3)
\psbezier[linewidth=0.01cm, linecolor=blue](1.33,2.6)(2.5,2.4)(2.5,2.4)(3.67, 2.6)
\psbezier[linestyle=dashed,linewidth=0.01cm, linecolor=blue](1.33,2.6)(2.5,2.8)(2.5,2.8)(3.67, 2.6)
\psbezier[linecolor=black](0, 3)(2,1.3)(3,1.3)(5, 3)
\psbezier[linecolor=black](0.75, 3)(2,2)(3,2)(4.25, 3)
\psline(-0.5, 3)(5.5,3)

\rput[l](3.9,3.2)%
	 {\rnode{node1}{{\blue $\Sigma$}}}
\rput[l](5.7,3)%
	 {\rnode{node1}{{\black $I$}}}
\tiny
\rput[l](-0.15,3.1)%
	 {\rnode{node1}{{\black $N_{0}$}}}
\rput[l](0.17,3.1)%
	 {\rnode{node1}{{\black $N$}}}
\rput[l](0.65,3.1)%
	 {\rnode{node1}{{\black $N_{I}$}}}
\psdots*(2.5,1.98)
\tiny
\rput[c](2.5,2.1)%
	 {\rnode{node1}{$p$}}

\end{pspicture}}
\caption{The family of deformations $\{N_{r}\}$, all anchored at Ext$(\Sigma)\cap I$. For $^{\Sigma}K_{\mu\nu}v^{\mu}v^{\nu}\leq 0$, $N_{I}$ and $N$ do not exist.}
\label{def}
\end{center}

\end{figure}

Suppose there exists a surface $N_{I}$ in $\{N_{r}\}$ such that $N_{I}\cap \text{Int}(\Sigma)\neq \emptyset$. Then, since $N_{I}$ is linked to $N_{0}$ via a series of smooth deformations, there exists a ``midway'' surface $N$ in the family of surfaces which coincides with $\Sigma$ at a set of points $\{p_{i}\}$ and is tangent to $\Sigma$ at those points. Fig. \ref{def} illustrates the deformation family. We focus on one coincident point $p$ of $\Sigma$ and $N$. 

As explained in \cite{Wall}, $^{\Sigma}\!K_{\mu\nu}v^{\mu} v^{\nu}$ is a measure of how much $\Sigma$ curves away from the tangent plane normal to $^{\Sigma}\!k^{\mu}$ with motion away from $p$ in the $v^{\mu}$ direction, where $^{\Sigma}\!k^{\mu}$ is the normal to $\Sigma$ when $\Sigma$ is not null, and we take $^{\Sigma}\!k_{\mu}$ to be the null generator for null $\Sigma$. Similarly, $^{N}\!K^{\rho}_{\mu\nu}v^{\mu}v^{\nu}\ ^{\Sigma}\!k_{\rho}$ measures how much $N$ curves away from the plane normal to $\Sigma$ at $p$ as one moves in the $v^{\mu}$ direction away from $p$ (restricted to motion tangent to $N$). $N$ is outside of $\Sigma$, so $\Sigma$ curves away from its tangent plane at least as much as $N$ does:
\begin{equation}^{\Sigma}\!K_{\mu\nu} v^{\mu}v^{\nu} \geq \phantom{} ^{N}\!K^{\rho}_{\mu\nu}v^{\mu}v^{\nu}  \phantom{}^{\Sigma}\!k_{\rho}\label{2.1}\end{equation} 
In particular:
\begin{align} & ^{\Sigma}\!K_{\mu\nu} h^{\mu\nu} \geq\phantom{}  ^{N}\!K_{\mu\nu}^{\rho} h^{\mu\nu} \ ^{\Sigma}\!k_{\rho}=0 \label{FirstCond}\\
&  ^{\Sigma}\!K_{\mu\nu} h^{\mu\nu} \geq 0 \label{negcond}\end{align}
\noindent where $h_{\mu\nu}$ is the induced metric on $N$. The equality in Eq. \ref{FirstCond} follows from the fact that $N$ is extremal.

$N$ is spacelike, so $h_{\mu\nu}$ is Riemannian and has no temporal components, which implies that $^{\Sigma}\!K_{\mu\nu}h^{\mu\nu}< 0$, since $^{\Sigma}\!K_{\mu\nu }v^{\mu}v^{\nu}< 0$. We therefore have a contradiction with Eq. \ref{negcond}, so the surfaces in the family $\{N_{r}\}$ never reach $\Sigma$, which in turn implies that $N_{I}$ does not exist. Moreover, it is impossible for $N$ to approach arbitrarily close to $\Sigma$, because in the limit Eq. \ref{negcond} would still hold, and we would again arrive at a contradiction.\footnote{If we are only interested in barriers for extremal surfaces with a particular dimensionality, Eq. \ref{FirstCond} may allow us to prove the existence of a barrier with a weaker condition than in Theorem 2.1.  For example, to prove that a spacelike codimension 1 surface is a barrier for $n$-dimensional extremal surfaces, it is sufficient if the sum of the largest $n$ eigenvalues of $^{\Sigma}\!K_{\mu\nu}$ are negative.  However, the analysis is more complicated for timelike or null barriers.}
\end{proof}

\subsection{So are $K = 0$ Surfaces}

\textbf{Theorem 2.2:} Let $\Sigma$ be a splitting surface as above, but we now take it to be totally geodesic, i.e. $^{\Sigma}\!K_{\mu\nu}=0$. Any $\Sigma$-deformable spacelike extremal surface which is anchored in the intersection of the boundary $I$ with one component of $M$ (as divided by $\Sigma$) remains in that component\footnote{Suppose that $M$ has some nontrivial topology or geometrical defect that creates an obstruction for $\Sigma$-deformability. If we can find another spacetime $M'$ which agrees with $M$ everywhere inside $\Sigma$ and also obeys the following conditions: (a) the extremal surface $N$ is the same in $M$ and $M'$, (b) the splitting surface $\Sigma$ is the same in $M$ and $M'$, and (c) $N$ is $\Sigma$-deformable in $M'$, then $\Sigma$ is still a barrier in $M$, by the proof of Theorem 2.2.}.

Note that in the case where $\Sigma$ is totally geodesic, it provides a barrier both for surfaces in Int$(\Sigma)$ and Ext$(\Sigma)$.

\begin{proof} 
The proof is identical for $N\subset \text{Int}(\Sigma)$ and $N\subset\text{Ext}(\Sigma)$. For concreteness we assume $N\subset \text{Ext}(\Sigma)$.

We take $N$ to be the midway surface as above, and we show that if $N$ agrees with $\Sigma$ at $p$ and is tangent to it, then $N\subset \Sigma$. This would imply that $N$ is no longer anchored at $I\cap \text{Ext}(\Sigma)$, and therefore it is impossible to deform a surface to reach and cross $\Sigma$ while maintaining boundary conditions in Ext$(\Sigma)$. The saturated equation is:
\begin{equation}0=\phantom{ }^{\Sigma}\!K=  \phantom{ }^{N}\!K^{\rho}\ ^{\Sigma}\!k_{\rho} =0\end{equation}
\noindent This shows that the two surfaces curve away in the normal direction equally, and is an indication that the two surfaces must agree everywhere on $N$ if they coincide at a point. We now show this in more detail.
\begin{figure}
\begin{center}
\psscalebox{1.7}
{
\begin{pspicture}(0,0)(4,4)

\psbezier(0,3)(1,1.65)(2,1.65)(3,3)
\psbezier(0, 1)(1,2.3)(2,2.3)(3, 1)
\psbezier[linewidth=0.01cm](0.33,2.6)(1.5,2.4)(1.5,2.4)(2.67, 2.6)
\psbezier[linestyle=dashed,linewidth=0.01cm](0.33,2.6)(1.5,2.8)(1.5,2.8)(2.67, 2.6)

\rput[l](2.9,3.2)%
	 {\rnode{node1}{$\Sigma$}}
\rput[l](2.9,0.8)%
	 {\rnode{node1}{$N$}}
\psdots*(1.5,1.98)
\tiny
\rput[c](1.5,2.1)%
	 {\rnode{node1}{$p$}}

\psline[linewidth=0.01cm](1.2, 2)(1.8,2)
\psline[linewidth=0.01cm](1.2, 1.95)(1.8,1.95)
\psline[linewidth=0.01cm](1.2, 1.9)(1.8,1.9)
\psline[linewidth=0.01cm](1.2, 1.85)(1.8,1.85)
\psline[linewidth=0.01cm](1.2, 1.8)(1.8,1.8)

\psline[linewidth=0.01cm](1.85, 2.1)(2.15,2.1)
\psline[linewidth=0.01cm](1.85, 2.05)(2.15,2.05)
\psline[linewidth=0.01cm](1.85, 2)(2.15,2)
\psline[linewidth=0.01cm](1.85, 1.95)(2.15,1.95)
\psline[linewidth=0.01cm](1.85, 1.9)(2.15,1.9)

\rput[c](1.5,1.6)%
	 {\rnode{node1}{$^{\Sigma}\!k_{\mu}\!(\!p\!)$}}
\psdots*(2,1.85)
\rput[c](2,1.7)%
	 {\rnode{node1}{$q$}}
\psdots*(2,2.11)
\rput[c](2,2.25)%
	 {\rnode{node1}{$s$}}
\rput[c](2.55,2)%
	 {\rnode{node1}{$^{\Sigma}\!k_{\mu}\!(\!s\!)$}}

\end{pspicture}}
\caption{A zoom-in near a neighborhood where 
$\Sigma$ and $N$ coincide. The horizontal lines at the point $p$ represent the covector $^{\Sigma}\!k_{\mu}(p)=\ ^{\Sigma}\!k^{\nu}(p)g_{\mu\nu}$, where $^{\Sigma}k^{\nu}(p)$ is the null generator for null $\Sigma$, or the normal for timelike or spacelike $\Sigma$. The horizontal lines at $s$ represent the covector $^{\Sigma}\!k_{\mu}(s)$, which is obtained by parallel transporting $^{\Sigma}\!k_{\mu}(p)$ along $\Sigma$.
}
\label{zoom}
\end{center}
\end{figure}
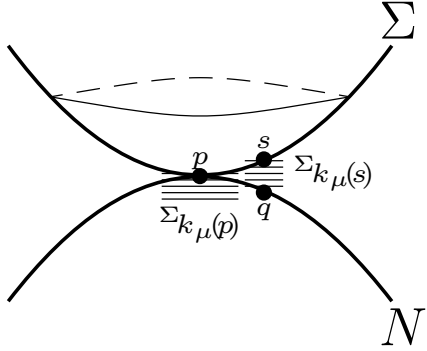

Suppose $\Sigma$ is null, and let $^{\Sigma}\!k^{\mu}$ be the null generator of $\Sigma$ directed towards $N$. The corresponding covector is defined: $^{\Sigma}\!k_{\mu} = g_{\mu\nu}\phantom{}^{\Sigma}\!k^{\nu}$, as illustrated in Fig. \ref{zoom}. Consider a small neighborhood $U$ of $p$ of length scale $\epsilon$. Let $q\in N\cap U$ and let $s\in \Sigma\cap U$. Define the coordinate $y(q)$:
\begin{equation} y(q) = \phantom{} ^{\Sigma}\!k_{\mu}(s)\left (s^{\mu}-q^{\mu}\right)\end{equation}
\noindent where $^{\Sigma}\!k_{\mu}(s)$ is obtained by parallel transporting $^{\Sigma}\!k_{\mu}(p)$ along $\Sigma$, and $(s^{\mu}-q^{\mu})$ is a vector pointing from $s$ to $p$ in the small neighborhood around $p$. Note that because $\Sigma$ is totally geodesic, $^{\Sigma}k_{\mu}(s)$ is null. Within $U$, we can always find an $s\in\Sigma$ such that $y(q)\ll\epsilon$. For such a choice of $s$, we therefore obtain:
\begin{equation}^{\Sigma}\!k_{\mu}(s)-\phantom{} ^{N}\!k_{\mu}(q)=\nabla_{\mu}y(q)+\mathcal{O}\left( \left(\nabla y\right)^{2}\right)\end{equation}
\noindent where $^{N}\!k_{\mu}(q)$ is obtained by parallel transporting $^{N}\!k_{\mu}(p)=\phantom{} ^{\Sigma}\!k_{\mu}(p)$ along $N$.

The null extrinsic curvatures therefore obey:
\begin{align} &^{\Sigma}\!K_{\mu\nu}-\phantom{} ^{N} \! K^{\rho}_{\mu\nu}\phantom{} ^{\Sigma}\!k_{\rho} = \nabla_{\mu}\nabla_{\nu}y \label{null}\\
&  ^{N}\!K^{\rho}_{\mu\nu}\phantom{} ^{\Sigma}\!k_{\rho} =-  \nabla_{\mu}\nabla_{\nu}y\\
&  ^{N}\!K^{\rho}\ ^{\Sigma}\!k_{\rho} =-  \nabla^{2}y\\
& 0=\nabla^{2}y\end{align}
\noindent Let $r$, defined on $\Sigma$, be the proper distance from $p$. We can use $r$ as a coordinate labeling points on $\Sigma$. Let $d\sigma$ be the volume element of a constant $r$ slice. Define $G(x)$ to be a Green's function on the ball of points $x$ with $r<R$, where
\begin{equation} - \nabla^{2}G(x) = \delta^{D-2}(x) \ ; \ \left. G\right |_{r=R}=0\end{equation}
\noindent where $\nabla$ is defined on $\Sigma$. For sufficiently small $R$, the metric on $\Sigma$ is approximately a flat Euclidean-signature metric, so $G\propto (r^{D-4}-R^{D-4})/(D-4)$ or $\ln(R/r)$ in $D=4$. Since $G(x)>0$ (regardless of what $D$ is), we find $\partial_{r}\left . G\right |_{r=R}<0$. When $R$ is small, these inequalities must continue to hold if the metric is slightly deformed by nonzero curvature. Integrating over a ball $B$ of radius $r=R$:
\begin{equation} 0 = \int\limits_{B} \nabla^{2}y\ G\ dr\ d\sigma = \int\limits_{\partial_{B}}\! y\  \partial_{r}G\ d\sigma\end{equation}
\noindent Since $y$ is never negative (that would contradict the assumption that $N$ does not cross $\Sigma$), we conclude that $y=0$ everywhere on a neighborhood of $p$. But this implies that $y=0$ everywhere, since it solves an elliptic equation, and thus $N\subset \Sigma$. This completes the proof for null $\Sigma$. This reasoning applies directly to the cases where $\Sigma$ is timelike or spacelike by simply substituting the $N$-pointing spacelike or timelike normal of $\Sigma$, respectively, for the null generator in the $N$ direction.  

\end{proof}


This proof allows us to generalize a previously-known result for totally geodesic surfaces:\\

\textbf{Corollary 2.3:} Let $\Sigma$ be a totally geodesic codimension $m$ surface in $M$. Let $N$ be a codimension $n\leq m$ spacelike extremal surface satisfying the following conditions:
\begin{enumerate} 
	\item $N$ coincides with $\Sigma$ at some point(s) $p$,
	\item $N$ does not cross $\Sigma$ anywhere (and therefore $N$ is tangent to $\Sigma$ at $p$),
\end{enumerate}
then $N\subset \Sigma$. \\

\begin{proof} Let $^{\Sigma}\! k^{\mu \ i}$ be the outwards-pointing normals to $\Sigma$ as above (null generators in the $N$ direction, if $\Sigma$ is null), where $i$ runs from $1,\cdots, m$. Let $^{\Sigma}\!k_{\mu}^{ i} = \phantom{} ^{\Sigma}\!k^{\nu \ i}g_{\mu\nu}$. Let $U$ be a neighborhood of $p$. We define $y(q)$ for $q\in N\cap U$ as above:
\begin{equation} y^{i}(q) = \phantom{}^{\Sigma}\! k_{\mu} ^{i}(s)\left ( s^{\mu} -q^{\mu}\right)\end{equation} 
\noindent where $s\in \Sigma\cap U$ is chosen so that $y^{i}(q)\ll\epsilon$. Then:
\begin{equation} ^{\Sigma}\! k_{\mu}^{i}(s) -\phantom{} ^{N}\!k_{\mu}^{i}(q) = \nabla_{\mu} y(q)^{i} + \mathcal{O}\left(\left (\nabla y\right)^{2}\right)\end{equation}
\noindent where for the case of null $\Sigma$, $^{\Sigma}\! k_{\mu}^{i}(p)=\phantom{} ^{N}\!k_{\mu}^{i}(p)$, and for timelike and spacelike cases, we just require them to be equal. As before, we obtain $^{\Sigma}\!k_{\mu}^{i}(s)$ by parallel transporting along $\Sigma$ and $^{N}\!k_{\mu}^{i}(q)$ by parallel transporting along $N$. The remaining reasoning in the proof of Theorem 2.2 above yields $y^{i}=0$, and we are done. \end{proof}

Note that Theorems 2.1 and 2.2, and Corollary 2.3 together also give a complete description of the behavior of extremal surfaces in the presence of barriers, provided that all disconnected components are $\Sigma$-deformable. Corollary 2.3 also leads to a natural corollary of Theorems 2.1 and 2.2, which combines the assumptions made in those theorems:

\textbf{Corollary 2.4:} Let $\Sigma$ be a splitting surface in $M$ such that $^{\Sigma}\!K_{\mu\nu}v_{i}^{\mu}v_{i}^{\nu}=0$ for some vector fields $\{v_{i}\}$ on $\Sigma$ 
and  $^{\Sigma}K_{\mu\nu}u_{i}^{\mu}u_{i}^{\nu}<0$ for all other vector fields $\{u_{i}\}$ on $\Sigma$. Then any $\Sigma$-deformable spacelike extremal surface which is anchored in Ext$(\Sigma)$ remains in Ext$(\Sigma)$. Moreover, although such extremal surfaces might conceivably come arbitrarily close to $\Sigma$ if they only propagate along the $v_{i}$ directions, they cannot approach $\Sigma$ if they propagate along the $u_{i}$ directions.
\begin{proof} As above, let $\{N_{r}\}$ be a family of $\Sigma$-deformable surfaces, all deformable from $N_{0}\subset \text{Ext}(\Sigma)$, and let $N$ be the midway surface between $N_{I}$ and $N_{0}$. We again focus on a coincident point $p\in \Sigma\cap N$. By Eq. \ref{negcond}, we have $^{\Sigma}\!K_{\mu\nu}h^{\mu\nu}\geq 0$, where $h_{\mu\nu}$ is the induced metric on $N$. We can decompose the metric into the components that lie along the $v$ directions and the $u$ directions:
\begin{equation*} h_{\mu\nu} = h^{(1)}_{\mu\nu} + h^{(2)}_{\mu\nu}\end{equation*}
\noindent where $h^{(1)}_{\mu\nu} = \sum\limits_{i} a_{i} v_{i\ \mu}v_{i \ \nu}$ and $h^{(2)}_{\mu\nu} = \sum\limits_{i} b_{i} u_{i\ \mu}u_{i \ \nu}$ for some constants $a_{i}$ and $b_{i}$. Eq. \ref{negcond} can be decomposed as follows:
\begin{equation} ^{\Sigma}\! K_{\mu\nu}h^{\mu\nu} = \phantom{}^{\Sigma}\!K_{\mu\nu}h^{(1)}\phantom{}^{\mu\nu}+ \phantom{}^{\Sigma}\!K_{\mu\nu}h^{(2)}\phantom{}^{\mu\nu}\label{decomposition}\end{equation}
\noindent If $N$ has no components in the $u_{i}$ directions, then $h^{(2)}\phantom{}^{\mu\nu}=0$, and Eq. \ref{decomposition} reduces to $^{\Sigma}\!K_{\mu\nu}h^{\mu\nu} = \phantom{}^{\Sigma}\!K_{\mu\nu}h^{(2)}\phantom{}^{\mu\nu} =0$. This is simply the case of Theorem 2.2, so $N$ could potentially get arbitrarily close to $\Sigma$ while remaining on Ext$(\Sigma)$.

If $N$ has at least one component in $u_{i}$ directions, then Eq. \ref{decomposition} yields $^{\Sigma}\!K_{\mu\nu}h^{\mu\nu} =\phantom{}^{\Sigma}\!K_{\mu\nu}h^{(1)}\phantom{}^{\mu\nu} + \phantom{}^{\Sigma}\!K_{\mu\nu}h^{(2)}\phantom{}^{\mu\nu} =\phantom{}^{\Sigma}\!K_{\mu\nu}h^{(2)}\phantom{}^{\mu\nu} <0$. This is simply the case of Theorem 2.1, so $N$ cannot approach $\Sigma$, and must always remain in Ext$(\Sigma)$. 

We conclude that if $N$ propagates in the directions in which $\Sigma$ has vanishing extrinsic curvature, $N$ can approach $\Sigma$, but if $N$ propagates in the directions in which $\Sigma$ has negative extrinsic curvature, $N$ cannot approach $\Sigma$. \end{proof} 

\subsection{Trapped Surface Barriers}

One consequence of Corollary 2.4 is that any null splitting surface $\Sigma$ which is foliated by surfaces where \emph{all} components of the null extrinsic curvature are nonpositive is an extremal surface barrier.  But if we are only interested in whether or not $\Sigma$ is a barrier to codimension 2 extremal surfaces, it turns out the we can do better: it is sufficient if the expansion $\theta \le 0$, i.e. the \emph{trace} of the null extrinsic curvature is nonpositive.  This is because when a codimension 2 extremal surface $N$ touches (but does not cross) $\Sigma$, we can take the trace of Eq. \ref{2.1} over all $D - 2$ spacelike directions to obtain:
\begin{equation}0 \ge \theta_\Sigma \ge \theta_N = 0.\end{equation}\label{thetas}
As in the case of Corollary 2.3, the above equation can only be saturated if $N$ lies on $\Sigma$.  This means that $\Sigma$ is a barrier to $\Sigma$-deformable spacelike extremal codimension 2 surfaces.

If we assume the null curvature condition $R_{\mu \nu} k^\mu k^\nu \le 0$, we can additionally show how to use a codimension 2 extremal surface to construct a barrier to other codimension 2 surfaces.

\textbf{Theorem 2.5:} Let $X$ be a codimension 2 spacelike extremal surface.  Let $\Sigma$ be the union of null congruences shot outwards from $X$ towards both the future and the past directions.  Assuming the null curvature condition, $\Sigma$ is a barrier to $\Sigma$-deformable codimension 2 surfaces anchored to Ext($\Sigma$).

\begin{proof}
By the standard construction, the Raychaudhuri equation implies that the two null congruences shot out from $X$ converge, so that $\theta_\Sigma \le 0$, where $\theta$ is defined as the expansion moving outwards away from $X$.  As in the proof of Theorems 2.1 and 2.2, we assume for contradiction that $N$ is a midway surface touching $\Sigma$ but not crossing it.  (It does not matter which of the two null congruences $N$ touches first, but whichever one it touches first, we are only interested in the $\theta$ of that congruence.) At the point of coincidence, Eq. \ref{thetas} implies that $\theta_\Sigma$ and $\theta_N$ both vanish.  Therefore, by the same reasoning as in Corollary 2.3, $N$ must lie entirely on $\Sigma$.  But then it cannot be anchored at Ext($\Sigma$).
\end{proof}

Even without assuming the null curvature condition, the following Corollary follows immediately:

\textbf{Corollary 2.6:} If $\Sigma$ is a null surface foliated by (marginally) outer trapped surfaces (i.e. $\theta \le 0$), then $\Sigma$ is a barrier to $\Sigma$-deformable codimension 2 surfaces anchored to Ext($\Sigma$).

Thus, if we are only interested in codimension 2 extremal surfaces (e.g. for purposes of holographic entanglement entropy \cite{RyuTakayanagi, HubenyRangamaniTakayanagi}), we can prove the existence of barriers in more circumstances.

\subsection{Outermost Barriers have a $K \geq 0$ Direction}

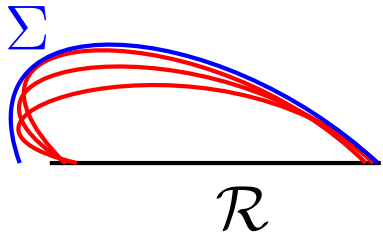
\begin{figure}[ht]
\begin{center}
\psscalebox{2}
{
\begin{pspicture}(0,0)(3,2)
\psline(0.9,1)(3.1,1)
\psbezier[linecolor=red](3,1)(2,2)(0,2)(1,1)
\psbezier[linecolor=red](3.05,1)(2,2)(-0.1,1.7)(1.01,1)
\psbezier[linecolor=red](3.08,1)(2,2)(-0.2,1.3)(1.08,1)
\psbezier[linecolor=blue](3.08, 1)(2,2)(0.3,2.1)(0.7,1)

\rput[l](2,0.7)%
	 {\rnode{node1}{$\mathcal{R}$}}
\blue{
\rput[l](0.6,1.9)%
	 {\rnode{node1}{$\Sigma$}}}

\end{pspicture}}
\caption{A illustration of a possible situation in which the extremal surface (in red) are anchored at $\mathcal{R}$, but they give rise to an outermost barrier $\Sigma$ (in blue above) whose exterior contains $\mathcal{R}$ as a proper subset.
}
\label{outermost}
\end{center}
\end{figure}

We proved that the presence of the surface $\Sigma$ with nonpositive spatial extrinsic curvature is a sufficient condition for an extremal surface barrier. This raises the question of whether this barrier is the closest one to the extremal surfaces. Consider some region on the boundary of the spacetime, $\mathcal{R}\subset I$. Let $X_{i}$ denote the spacelike extremal surfaces anchored at $\mathcal{R}$. We define the outermost barrier $\Sigma$ for surfaces anchored at $\mathcal{R}$ to be given by :
\begin{equation*}\Sigma = \partial \left (\bigcup\limits_{i} X_{i} \right).\end{equation*}
(For some (non-AlAdS) spacetimes, $\mathcal{R}$ may be a proper subset of Ext$(\Sigma)$, as shown in Fig. \ref{outermost}.)   The above definition is partly motivated by \cite{CzechKarczmarekNogueiraVanRaamsdonk}, although the case under discussion there involved only minimal area surfaces, whereas we consider all extremal surfaces. Note that we can similarly define the outermost barrier for a subset $\{Y_{i}\}$ of all spacelike extremal surfaces anchored at $\mathcal{R}$ by defining $\Sigma_{Y} = \partial \left (\bigcup\limits_{i}Y_{i} \right)$.

We now prove some properties of the outermost barrier.\\

\textbf{Theorem 2.7:} Let $\Sigma$ be an outermost barrier in $M$, and let $\{N_{r}\}$ be a family of extremal surfaces that can all be smoothly deformed from some initial surface $N_{0}\subset \text{Ext}(\Sigma)$ such that some surface $N\in \{N_{r}\}$ is arbitrarily close to $\Sigma$. Then $^{\Sigma}\!K_{\mu\nu}h^{\mu\nu}\geq 0$, where $h^{\mu\nu}$ is the induced metric on $N$.
\begin{proof}
By definition of outermost barrier, for any point $p$, spacelike extremal surfaces in Ext$(\Sigma)$ can be deformed to either come arbitrarily close to $\Sigma$ or coincide with and be tangent to $\Sigma$ at $p$. In the latter case, we can simply take the limiting surface that touches $\Sigma$. In any small neighborhood where $\Sigma$ and one of these spacelike extremal surfaces touch or nearly touch, Eq. \ref{negcond} yields $^{\Sigma}\!K_{\mu\nu}h^{\mu\nu}\geq 0$, where $h_{\mu\nu}$ is the induced metric on the extremal surface, and we are done. \end{proof}

In particular, the extremal surface is spacelike and $h_{\mu\nu}$ is Riemannian, so in order for all of the components of $^{\Sigma}\!K_{\mu\nu}$ to add up to a nonnegative number, at least one eigenvalue of $^{\Sigma}\!K_{\mu}^{\nu}$ must be nonnegative.

Whenever we can find a barrier of entirely negative extrinsic curvature, we can therefore find a barrier of partly nonnegative curvature.  This is in agreement with Theorem 2.1, which implies that extremal surfaces cannot come arbitrarily close to a surface with strictly negative curvature. The existence of a barrier is therefore quite generic, since we expect it to occur for any surface $\Sigma$ which splits the spacetime into 2 regions and has negative or vanishing extrinsic curvature components.  


Many of the known analytic solutions that are AlAdS admit a totally geodesic surface that acts an as extremal surface barrier, as well as multiple surfaces with negative extrinsic curvature. Totally geodesic surfaces are quite special and therefore spacetimes containing them are not representative of general AlAdS spacetimes. We expect, however, that small perturbations of spacetimes that admit splitting surfaces with negative extrinsic curvature do again result in spacetimes with splitting surfaces of nonpositive extrinsic curvature. The existence of such surfaces is thus stable under small perturbations of the metric. 

We therefore conclude that \textit{many generic AlAdS spacetimes admit an extremal surface barrier}. Moreover, we show in the examples below that the barrier often separates a region of proximity to a singularity from the boundary of the spacetime. In Section \ref{Singularities}, we further prove that for certain spacetimes, the existence of a compact outermost barrier implies the existence of singularities.

The implications of this barrier in the context of AdS/CFT are disturbing: as rule of thumb, the best-understood aspect of the duality dictionary is generically limited in the scope of information it contains about the bulk. This at best suggests that we must change our approach towards extracting bulk information by using probes that are not extremal surfaces. At worst indicates that complete information of a large class of bulk geometries may simply not be contained in the dual field theory \cite{MarolfWall}.

\section{Examples} \label{examples}
\noindent In this section, we provide some examples of spacetimes with barriers.  Some of these barriers are totally geodesic, others have negative extrinsic curvature.  The latter condition is stable under small perturbations, although the former is not.


\subsection{Pure AdS}

\noindent The simplest example is pure AdS itself. If we consider the metric on the Poincar\'e patch:
\begin{equation} ds^{2}= \frac{1}{z^{2}}\left ( -dt^{2} + dx_{i}^{2} +dz^{2}\right)\end{equation}
\noindent Surfaces of constant $t$ or surfaces of constant $x_{i}$ are totally geodesic. Since extremal surfaces in pure AdS remain on constant time slices, it is clear that they do not cross constant $t$ totally geodesic surfaces. Similarly, a surface which is anchored between $x_{i}=-x_{0}$ and $x_{i}=x_{0}$ will not propagate to $x_{i}>x_{0}$ or $x_{i}<-x_{0}$. 

\subsection{AdS Cosmology}\label{cosmology}
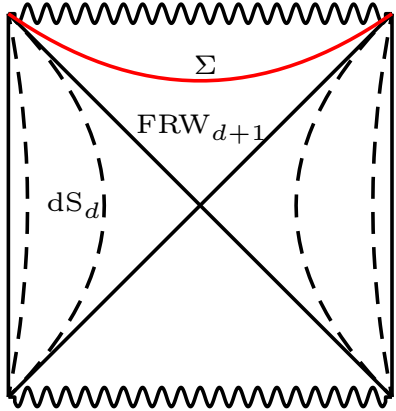
\begin{figure}
\begin{center}
\psscalebox{1.7}
{\begin{pspicture}(0,0)(4,4)
\psline{}(0,0)(0,3)
\psline{}(3,0)(3,3)
\psline{}(0,0)(3,3)
\psline{}(3,0)(0,3)
 \psset{coilwidth=0.15}
\pscoil[coilaspect=0,coilarmA=0.05cm, coilarmB=0.05cm](0,0)(3,0)
\pscoil[coilaspect=0,coilarmA=0.05cm, coilarmB=0.05cm](0,3)(3,3)
\psset{dotsize=0.2 0}
\psbezier[linestyle=dashed](0,3)(1,2)(1,1)(0,0)
\psbezier[linestyle=dashed](0,3)(0.2,2)(0.2,1)(0,0)
\psbezier[linestyle=dashed](3,3)(2,2)(2,1)(3,0)
\psbezier[linestyle=dashed](3,3)(2.8, 2)(2.8,1)(3,0)
\psbezier[linecolor=red](0,3)(1,2.3)(2,2.3)(3,3)
\tiny
	\rput[l](1,2.1)%
	 {\rnode{node1}{FRW$_{d+1}$}}

\rput[l](1.45,2.6)%
	 {\rnode{node1}{$\Sigma$}}

\rput[l](0.3,1.5)%
	 {\rnode{node1}{dS$_{d}$}}
\end{pspicture}
}
\caption{The isotropic AdS$_{d+1}$ cosmology. $\Sigma$ is the surface where $a(t)$ is maximized.}
\label{frw}
\end{center}
\end{figure}

The next example we consider is an isotropic AdS cosmology. The metric on global AdS can be written, via a coordinate transformation (see e.g. \cite{MaldacenaPimentel}), as an open Friedmann-Robertson-Walker (FRW) universe in the interior of the lightcone, and foliated into de Sitter slices in the exterior of the lightcone:
\begin{align} & ds^{2}_{\text{int}} = -dt^{2} +\sin^{2}(t) \left (d\chi^{2} + \sinh^{2}\chi \ d\Omega^{2}\right)\\
& ds^{2}_{\text{ext}} = dr^{2} + \sinh^{2}(r)\left (-d\tau^{2} +\cosh^{2}\tau\  d\Omega^{2}\right)\end{align} 
\noindent where the two are related via an analytic continuation (see Fig. \ref{frw}). Any small perturbation to this metric results in a curvature singularity replacing the coordinate singularity. If we couple the metric to, say, a scalar field, as in \cite{HertogHorowitz}, the resulting geometry features a curvature singularity (the first indication of some sort of an accumulation surface for codimension 2 surfaces in this geometry was found in \cite{MaldacenaPimentel}):
\begin{align} & ds^{2}_{\text{int}} = -dt^{2} +a^{2}(t) \left (d\chi^{2} + \sinh^{2}\chi\  d\Omega^{2}\right)\\
& ds^{2}_{\text{ext}} = dr^{2} + b^{2}(r)\left (-d\tau^{2} +\cosh^{2}\tau \ d\Omega^{2}\right) \label{dS}\end{align} 
\noindent where $a(0)=0$ and $b(0)=1$ at the lightcone and $a(t_{\text{singularity}})=0$ is a big-crunch collapse. It was found in \cite{WIP} that this spacetime contains a totally geodesic barrier: there is a point $t_{m}$ at which $a(t_{m})$ reaches a maximum. The extrinsic curvature of any constant $t$ surface with respect to the past-pointing timelike normal is given by:
\begin{equation*} K_{ij}= a'(t) g_{ij}\end{equation*}
\noindent where $g_{ij}$ is the FRW metric inside the lightcone. The constant $t=t_{m}$ surface $\Sigma$ is therefore totally geodesic, and surfaces with $t>t_{m}$ have negative extrinsic curvature. We demonstrate the barrier at work here using spacelike radial geodesics as an example of extremal surface probes. This is easily generalized to extremal surfaces with fewer symmetries, but for brevity we limit ourselves to the simplest case. 

Consider a geodesic with endpoints at $\theta=0$ and $\theta=\pi$, where $\theta$ is one of the suppressed angles in Eq. \ref{dS}. The length functional for this geodesic within the FRW region is:
\begin{equation} \mathcal L = \int \sqrt{-dt^{2} +a^{2}(t) d\chi^{2}} = \int\sqrt{-t'(\chi)^{2} + a^{2}(t(\chi))}d\chi\end{equation}
\noindent where we have chosen $\chi$ to parametrize the geodesic. The boundary conditions $\theta=0$ and $\theta=\pi$ imply $t'(\chi=0)=0$. We solve for the geodesic from the lightcone interior outwards, starting from $\chi=0$ to the lightcone and the exterior region, using the reflection symmetry in $\theta$ along the geodesic path. 

Since $\chi$ is a cyclic coordinate, the Hamiltonian is conserved:
\begin{equation} \mathcal{H}(t) = - \frac{a^{2}(t)}{\left ( a^{2}(t) -t'(\chi)^{2}\right)^{1/2}} = \mathcal{H}\left(t(\chi=0)\right) = a\left(t(\chi=0)\right).\end{equation}
Solving for $t'(\chi)$ yields:
\begin{equation} t'(\chi) = \pm a\left(t(\chi)\right) \left ( 1- \left(\frac{a\left(t(\chi)\right)}{a\left(t(0)\right)}\right)^{2}\right)^{1/2}.\label{tprime}\end{equation}
Since the scale factor vanishes at the lightcone and at the singularity, and reaches its maximum at some intermediate time $t=t_{m}$, it follows from Eq. \ref{tprime} that any geodesic with $t(\chi=0)>t_{m}$ must propagate towards progressively larger $t$. Geodesics with $t(\chi=0)=t_{m}$ stay on $t(\chi)=t_{m}$ for all $\chi$. This is precisely what Theorem 2.2 states: extremal surfaces in the interior of $\Sigma$ (i.e. $t>t_{m}$) remain in the interior, extremal surfaces in the exterior of $\Sigma$ (i.e. $t<t_{m}$) remain in the exterior, and extremal surfaces that coincide with $\Sigma$ (i.e. $t=t_{m}$) lie on $\Sigma$, where $\Sigma$ is the totally geodesic surface given by $t=t_{m}$. 

One implication of this result is that, in the isotropic AdS Cosmology, only geodesics on the exterior of $\Sigma$ make it to the lightcone at $t=0$ and from there to the boundary. The totally geodesic surface prevents boundary-anchored probes from getting arbitrarily close to the singularity.

\subsection{Black Holes}

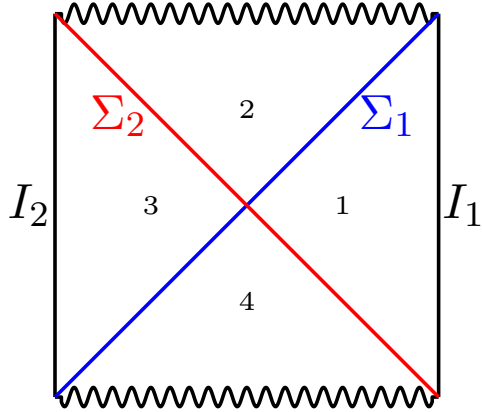
\begin{figure}
\begin{center}
\psscalebox{1.7}
{
\begin{pspicture}(-1,0)(4,4)

\psline{}(0,0)(0,3)
\psline{}(3,0)(3,3)

 \psset{coilwidth=0.15}
\pscoil[coilaspect=0,coilarmA=0.05cm, coilarmB=0.05cm](0,0)(3,0)
\pscoil[coilaspect=0,coilarmA=0.05cm, coilarmB=0.05cm](0,3)(3,3)
\psset{dotsize=0.2 0}

\tiny
	\rput[c](1.5,2.25)
	 {\rnode{node1}{2}}
\rput[c](2.25,1.5)
	 {\rnode{node1}{1}}
\rput[c](1.5,0.75)%
	 {\rnode{node1}{4}}
\rput[c](0.75,1.5)
	 {\rnode{node1}{3}}

\small
\rput[c](3.2,1.5)
	 {\rnode{node1}{$I_{1}$}}
\rput[c](-0.2,1.5)
	 {\rnode{node1}{$I_{2}$}}
\rput[c](2.6,2.2)
	 {\rnode{node1}{\blue{$\Sigma_{1}$}}}
\rput[c](0.5,2.2)
	 {\rnode{node1}{\red{$\Sigma_{2}$}}}

\psbezier[linecolor=blue](0,0)(3,3)(0,0)(3,3)
\psbezier[linecolor=red](3,0)(0,3)(3,0)(0,3)

\end{pspicture}}
\caption{The Schwarzschild-AdS$_{d+1}$ black hole. $\Sigma_{1}$ and $\Sigma_{2}$ are totally geodesic splitting surfaces, and are therefore by Theorem 2.2 extremal surface barriers.}
\label{BH}
\end{center}
\end{figure}


Another prominent example of a spacetime with a totally geodesic barrier surface as prescribed in Section \ref{theorem} includes any black hole with a stationary horizon \cite{Mars, Hajicek}\footnote{Stationary geons, which are not black holes, also exhibit a barrier at the horizon.}. This category includes AlAdS spacetimes such as Schwarzschild-AdS, RN-AdS, Kerr-AdS, the BTZ black hole, the planar AdS black hole (the horizon in this case was separately proven to be a barrier surface in \cite{Hubeny}), and the hyperbolic AdS black hole.  In particular, in AdS-Schwarzschild, there are two totally geodesic surfaces: the past- and future-directed horizons, denoted $\Sigma_{1}$ and $\Sigma_{2}$ in Fig. \ref{BH}. Spacelike extremal surfaces anchored at $I_{1}$ must therefore always remain in region 1, while those anchored at $I_{2}$ must remain in  region 3 of the spacetime. Extremal surfaces anchored at both $I_{1}$ and $I_{2}$ are not constrained by our theorems to remain in any part of the spacetime. In fact, \cite{FidkowskiHubenyKlebanShenker} 
found that spacelike geodesics with one endpoint at $I_{1}$ and one endpoint at $I_{2}$ can probe arbitrarily close to the singularity, and in particular, can cross both $\Sigma_{1}$ and $\Sigma_{2}$. We emphasize that this does not contradict our results in the previous section because these geodesics are anchored to two boundaries. As a point of interest, \cite{HartmanMaldacena} found a barrier surface, or accumulation surface, for codimension 2 extremal surfaces anchored to both boundaries, while, as noted above, radial geodesics observe no such barrier \footnote{The Vaidya spacetime, which we discuss in more detail in Sec. \ref{Discussion}, has just one boundary, does not have a barrier for all extremal surfaces because spacelike geodesics which enter at early times can leave at arbitrarily late times. However, there is an effective barrier for surfaces that are anchored to the boundary at late times, and there may be a barrier for codimension 2 surfaces. Following the first version of this paper, an analysis of the behavior of extremal surfaces in some Vaidya spacetimes was done in \cite{HubenyMaxfield}.}. 

It is clear at this point that spacetimes with singularities seem to have a particular proclivity for admitting barriers. This is not a coincidence: Corollary 2.6 states that null surfaces foliated by (marginally) trapped surfaces act as barriers at least for codimension 2 extremal surfaces.  If we additionally assume the stronger statement\footnote{If we assume spherical symmetry, this condition automatically follows for trapped surfaces.} that \emph{all} components of the null extrinsic curvature are nonpositive, we found in Corollary 2.4. that it is a barrier to extremal surfaces of every dimension.

In particular, when the null barrier $\Sigma$ is spatially compact, i.e. the entire boundary is in Ext$(\Sigma)$, then $\Sigma$ is ruled by trapped surfaces.  If the spacetime is furthermore globally (or AdS) hyperbolic, spatially noncompact, and obeys the null curvature condition, the existence of trapped surfaces guarantees the existence of singularities and horizons \cite{HawkingEllis}. 
In the next section, we will prove a partial converse of our conclusions from Corollary 2.4: barriers imply the existence of trapped surfaces or singularities, at least for spacetimes which admit a totally geodesic spacelike slice.

It is worthwhile here to comment on any bearing this might have on the recent controversy over the completeness of the AdS/CFT correspondence. The question of whether the black interior is fully described by the boundary field theory is particularly relevant to this discussion. In \cite{AMPSS, MarolfPolchinski, Bousso}, it was shown that there exist operators in the static black hole interior that cannot exist in the dual field theory. The fact that a well-used probe of the bulk geometry cannot reach into a bulk region which is not completely described by the boundary field theory may not be coincidental. Since the limited reach of extremal surfaces in the static black hole geometry is simply a special case of the barrier theorems presented in Section \ref{theorem}, we may expect that other black hole AlAdS geometries admitting barriers may manifest similar incompleteness on the dual field theory side. One may also speculate that field theories dual to other geometries, not necessarily black holes, may exhibit the same behavior near a singularity.

\section{Barriers and Singularities}\label{Singularities}
\noindent  The examples in the previous section suggest that there is some connection between singularities (particularly singularities masked by horizons), and extremal surface barriers.  In this section, we will prove this result in some special cases.  Consider a spacetime manifold $M$ containing a totally geodesic spacelike slice $S$.  Such slices can be found, for instance, in static spacetimes, or more generally, spacetimes with a moment of time reflection symmetry.  Since we intend to drop in extremal surfaces from the boundary of $S$, we must assume that $S$ is noncompact.  Because $S$ is totally geodesic, these extremal surfaces remain on $S$.

Suppose further that there is an outermost barrier $\Sigma$ such that the intersection between $\Sigma$ and $S$ is nonempty and compact.  Then we prove below that $S$ either admits a singularity, or else a marginally trapped surface.

If, in addition, we assume global hyperbolicity (
including AlAdS spacetimes with the appropriate generalization of global hyperbolicity), the null curvature condition ($R_{\mu\nu} k^{\mu} k^{\nu} \ge 0$ for all null vectors $k^{\mu}$) and the generic condition (roughly, that each null ray encounters at least a little bit of null curvature or shear) then the existence of a marginally trapped surface itself implies the existence of a singularity somewhere on $M$ \cite{HawkingEllis}.

We first prove a lemma:\\

\textbf{Theorem 4.1:} No compact extremal surface barrier $\Sigma$ in $M$ ever touches extremal surfaces anchored to $I$.

\begin{proof} 
We prove this by contradiction.  Let $\Sigma$ be a barrier, and suppose there exists a spacelike extremal surface $N$ which is anchored to $I$. By assumption, $ N\subset \text{Ext}(\Sigma)$ everywhere except where it coincides. 
In any neighborhood of such a coincident point $p$, there exists a point such that:
\begin{equation} ^{\Sigma}\!K_{\mu\nu}h^{\mu\nu} > \phantom{}^{N}K_{\mu\nu}^{\rho}h^{\mu\nu} \ ^{\Sigma}\!k_{\rho}=0\end{equation}
\noindent where $^{\Sigma}\!k_{\rho}$ is the null generator of $\Sigma$ for null $\Sigma$, and the $N$-directed normal to $\Sigma$ otherwise. 

$N$ cannot lie on $\Sigma$ for a continuous neighborhood (otherwise it would lie on $\Sigma$ everywhere, which would violate the boundary conditions on $N$).  $N$ can therefore touch $\Sigma$ while remaining anchored on the boundary.  But then we can slightly deform $N$ in a neighborhood of $p$, to make a new surface $N'$ which crosses $\Sigma$.  Using the elliptical equation of extremality, we can solve for $N'$ outside the neighborhood as well.  Because the perturbation is small, $N'$ must still be anchored to the boundary.  This shows that $\Sigma$ is not a barrier, and we have arrived at a contradiction.  \end{proof}


Theorem 4.1 may at first seem to indicate that compact outermost barriers cannot exist, since we have defined outermost barriers as the boundary of the union of some set of extremal surfaces: one might naturally ask how the outermost barrier is constructed, if extremal surfaces cannot touch it. The barrier must be constructed as a limit of extremal surfaces that come arbitrarily close to touching it.  We assume therefore, that one can find sequences of extremal surfaces approaching any point $p$, whose limit is an extremal surface tangent to $\Sigma$ at $p$.  However, the limiting extremal surface must not be anchored anymore to the boundary, or it would contradict Theorem 4.1.  We will make use of this limit construction of the outermost barrier in the proof of Theorem 4.2 below, where we prove a condition relating the existence of barriers to the presence of singularities and trapped surfaces.

Because we are taking a limit of extremal surfaces, one may worry that this limit will be related to the boundary in a bizarre way, perhaps by spiraling around, getting arbitrarily close to the boundary without actually being anchored to it.  We will deal with these pathological cases by including them in the following definition: a surface $N$ is \textit{weakly anchored} if there exists a $d\in\mathbb{R}$ such that all points $p\in N$ are less than distance $d$ away from the boundary $I$ along $N$ (after compactification of $I$). 

\begin{figure}
\begin{center}
\includegraphics[width=5cm]{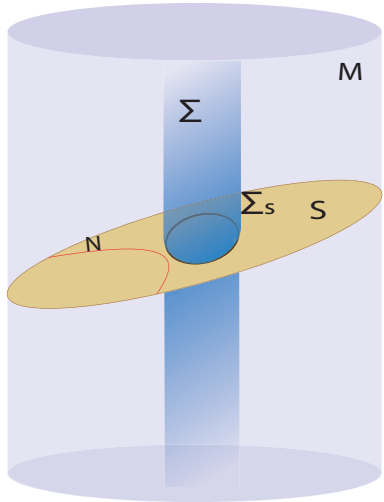} 
\caption{A spacetime $M$ with a timelike barrier $\Sigma$, projected onto a spacelike codimension 1 totally geodesic slice $S$ yields a compact barrier $\Sigma_{S}$ for extremal surfaces on $S$.}
\label{full}
\end{center}
\end{figure}

\textbf{Theorem 4.2:} Let $M$ admit a totally geodesic codimension 1 spacelike slice $S$ \footnote{We assume that all points on $S$ are a finite distance away from the boundary after compactification.}. Given a compact, nonempty, outermost barrier $\Sigma_{S}$ for weakly-anchored, codimension 2, $\Sigma_{S}$-deformable (i.e. where the deformations are restricted to $S$) spacelike extremal surfaces on $S$ (e.g. Fig. \ref{full}), then one of the following is true:
\begin{enumerate}
	\item There exists a singularity outside or on $\Sigma_{S}$, or
	\item $\Sigma_{S}$ is a marginally trapped surface (in either time direction).
\end{enumerate}

\begin{proof}
Since $\Sigma_{S}$ is an outermost barrier, there exist extremal surfaces that either coincide with it and are tangent to it at coincident points, or come arbitrarily close to coinciding with it. The former case is ruled out by Theorem 4.1. 

Let $\{N_{r}\}$ be a family of spacelike extremal surfaces that approach arbitrarily close to some point $q \in \Sigma_{S}$ while remaining weakly anchored on $I_{S}=I\cap S$.  In a neighborhood of $q$, we take the limit of these surfaces $N'$ of $\{N_{r}'\}$ that reaches $\Sigma$ at $q$.  

Since we are operating at the ``physics'' standard of rigor, we assume without proof that (a) this limit must exist for some sequence of extremal surfaces, and that (b) the resulting extremal surface may be extended outside the neighborhood of $q$ by solving the elliptical equation of extremality.  When we solve this equation, by Theorem 4.1, $N'$ must no longer be weakly anchored on $I_{S}$.


Since $N'$ cannot be weakly anchored to $I_{S}$, part of it must terminate somewhere in the interior of the spacetime. It can do it in one of the following two ways (Fig. \ref{Thm3}):
\begin{description}
	\item[(a)] $N'$ terminates in a singularity on or outside $\Sigma_{S}$, or
	\item[(b)] $N'$ spirals outside $\Sigma_{S}$ forever, without ever coming to an end.  In that case, we can define $N_{\infty}'$ to be the limit set of the extremal surface as it gets arbitrarily far away from the points where it is anchored to the boundary.\footnote{If $N'$ is not (weakly) anchored to the boundary anywhere, then $N_{\infty}' = N'$.} $N_{\infty}'$ must be bounded by some compact innermost surface $\Sigma_{S}'$, where $\Sigma_{S}'\cap I = \emptyset$. \footnote{Note that $N'_{\infty}$ cannot come arbitrarily close to the boundary, since then it would be boundary-anchored by the generalized definition above.}
\end{description}
If (a) is true, we get the first part of the claim. We now show that (b) results in the second part of the claim. To do so, we need to show that $\Sigma_{S}=\Sigma_{S}'$. We do this by contradiction.
\begin{center}
\begin{figure}[t]
\centering

\subfigure[ ]{
\centering
\includegraphics[width=0.3 \textwidth]{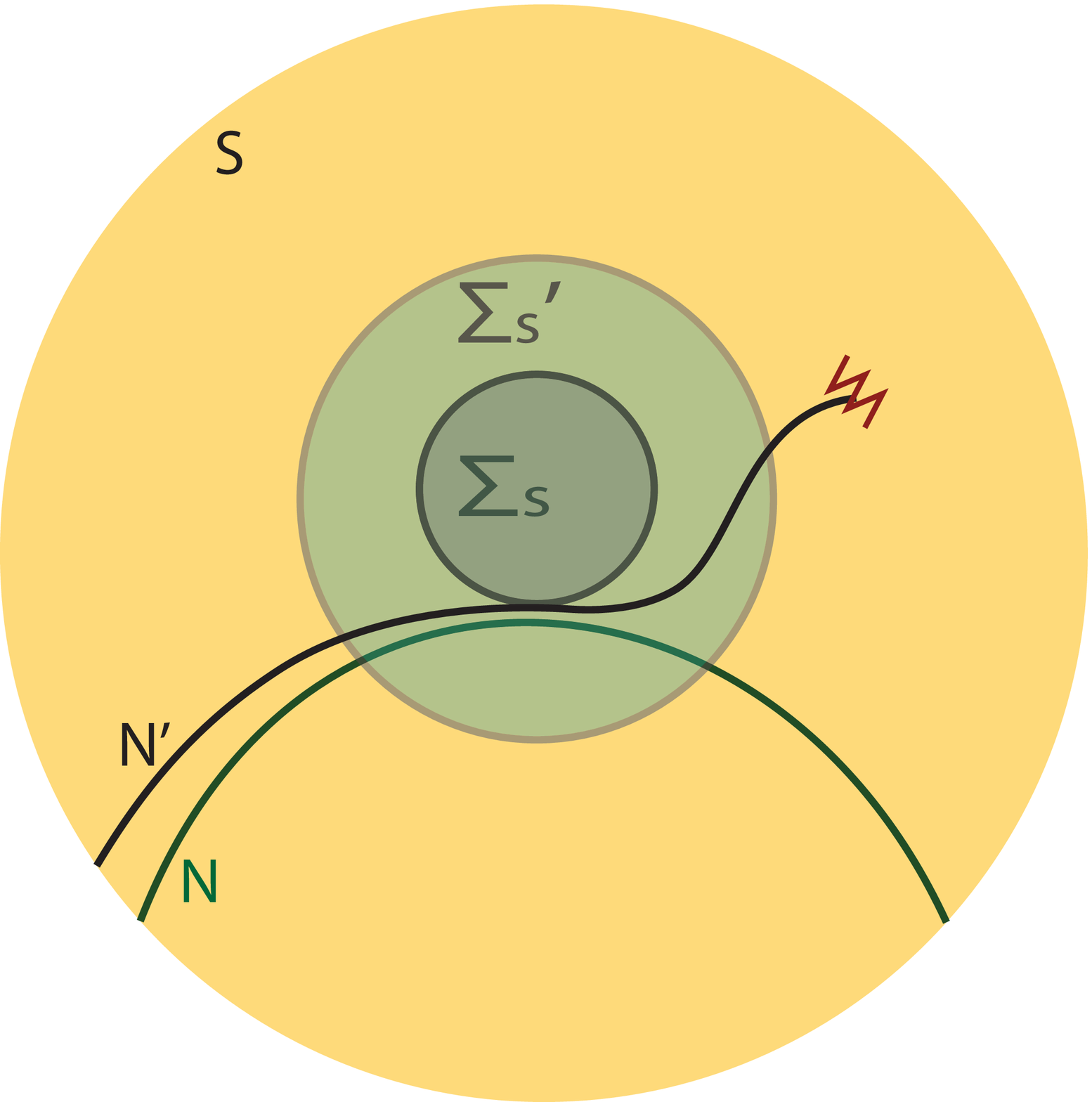}
}
\qquad
\subfigure[]{
\centering \includegraphics[width=0.3\textwidth]{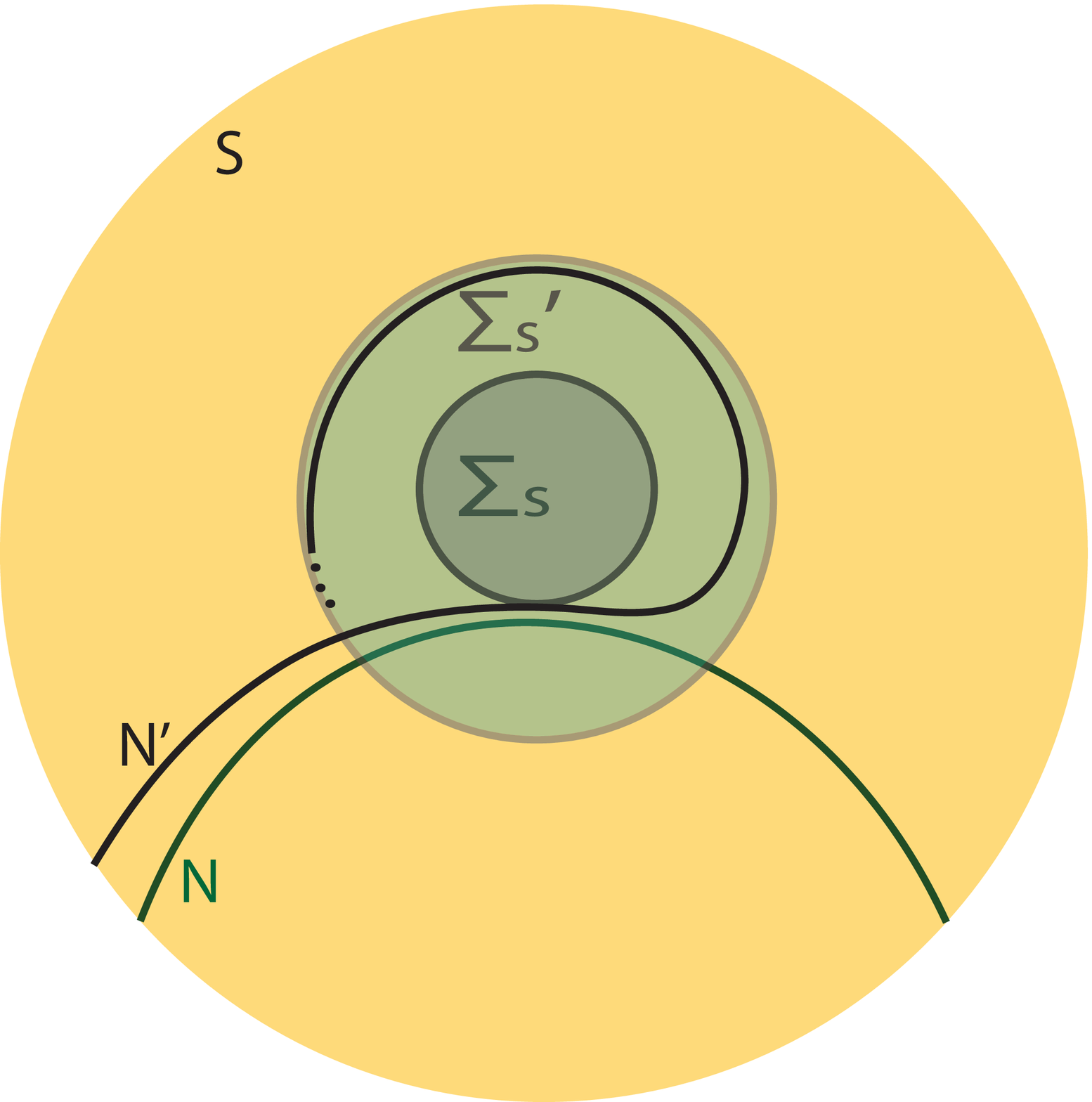}
}
\caption{ (a) The original extremal surface $N$ and its limiting surface $N'$, in the case where $N'$ terminates at a singularity, (b) The extremal surfaces $N$ and $N'$, where $N'$ spirals into some limiting surface bounded by $\Sigma_{S}'$.}
\label{Thm3}
\end{figure}
\end{center}

Suppose $\Sigma_{S}'\neq \Sigma_{S}$. Then $\Sigma_{S}'\subset\text{Ext}(\Sigma_{S})$. This means that it is \emph{not} a barrier for extremal surfaces weakly anchored at $I$, and therefore by Theorem 2.1 there must exist a point $p\in \Sigma_{S}'$ such that
\begin{equation} \text{tr}_{S}\left( \phantom{}^{\Sigma_{S}'}K_{\mu\nu}\right)|_{p}>0.\label{tracepos}\end{equation}

On the other hand, by construction, $N'_{\infty}\subset \text{Int}(\Sigma_{S}')$ can come arbitrarily close to $\Sigma_{S}'$ at any point $p\in\Sigma_{S}'$.  This implies that the following holds everywhere:
\begin{equation} \text{tr}_{S}\left( \phantom{}^{\Sigma_{S}'}K_{\mu\nu}\right)\leq0.\label{traceneg}\end{equation}
(The proof is essentially the same as Theorem 2.7, but reversing the roles of the interior and exterior.)

We have arrived at a contradiction, so instead we must assume that $\Sigma_{S}'=\Sigma_{S}$.  This tells us that $N'$ must in fact coincide with $\Sigma_{S}$.  Since $N'$ is extremal, it follows that $\Sigma_{S}$ satisfies
\begin{equation} \text{tr}_{S}\left( \phantom{}^{\Sigma_{S}}K_{\mu\nu}\right)=0.\end{equation}
This is the same as saying that $\Sigma_{S}$ is a marginally trapped surface, which proves the second part of the claim. \end{proof}

The assumption that there exists an outermost barrier $\Sigma_{S}$ for codimension 2 extremal surfaces living on $S$ is reasonable: if we take $M$ to have a totally geodesic slice $S$ and a compact outermost barrier $\Sigma$ such that $\Sigma\cap S\neq \emptyset$, then in general, we expect that $\Sigma\cap S$ will be compact, and be a barrier for extremal surfaces on $S$. There should then exist some outermost $\Sigma_{S}$ which, at least for AlAdS spacetimes, will still be compact.  (If there is a barrier for \emph{all} extremal surfaces, then obviously there must be a barrier for those which are codimension 2 and restricted to $S$.)

If we further assume global hyperbolicity, the null curvature condition, and the generic condition, we are able to say that marginally trapped surfaces only occur in geodesically incomplete spacetimes, so that barriers in these spacetimes exist only in the presence of singularities \cite{HawkingEllis}.  Singularities are therefore intrinsically linked to extremal surface barriers.  Assuming some form of cosmic censorship, these singularities must be hidden behind horizons, thus outermost barriers are also linked to horizons.

In particular, the direct application of Theorem 4.2 to the AdS/CFT correspondence suggests that in general, extremal surface probes are not good probes for spacetimes with singularities.

We expect that Theorem 4.2 can be further generalized to both the noncompact barrier case, as Section \ref{cosmology} suggests, and to any time-dependent geometry, although this may require assuming the null curvature condition (implied by the null energy condition and Einstein's equations), even to prove the existence of the trapped surface.\footnote{For example, the proof of strong subadditivity for extremal surfaces can be proven without the null curvature condition in the static case \cite{HeadrickTakayanagi}, but requires it in the dynamical case \cite{Wall2}.}
\section{Discussion}\label{Discussion}

\noindent We proved in Theorem 2.1, Theorem 2.2, and Corollary 2.4 that a splitting surface with nonpositive extrinsic curvature acts as a barrier to extremal surfaces (at least those which are $\Sigma$-deformable).  This can include spacelike, timelike, or null barrier surfaces.  If we are only interested in barriers to codimension 2 surfaces (as used in the holographic entanglement entropy conjecture \cite{RyuTakayanagi, HubenyRangamaniTakayanagi}), then we have shown in Theorem 2.5 and Corollary 2.6 that null surfaces foliated by (marginally) trapped surfaces are barriers.



Conversely, for spacetimes with a totally geodesic slice, the existence of a barrier guarantees the existence of either singularities or else trapped surfaces for spacetimes with a totally geodesic slice $S$, so long as the intersection of the barrier and $S$ is compact.  We also showed that outermost barriers have at least one nonnegative extrinsic curvature component (Theorem 2.7), and we argued that extremal surface barriers occur somewhat generically in AlAdS spacetimes.  This presents a setback to attempts to reconstruct the bulk geometries from field theory operators using extremal surface probes.

These results are particularly troubling in light of how pervasive extremal surface probes are in applications of the AdS/CFT correspondence.  If there are regions which cannot be reconstructed from the entire boundary, than either those regions do not really exist as semiclassical regions (e.g. because there is a firewall at their boundary), or else if they do exist, the information inside of them must be contained in a new factor of the Hilbert space which is in addition to the boundary \cite{MarolfWall}.

We have also studied cases in which there is a barrier to extremal surfaces located on only \emph{part} of the boundary.  In this case it is not surprising that the entire bulk cannot be reconstructed.  Thus the location of the barrier might give clues about how much of the bulk can be reconstructed from a CFT region.  See \cite{CzechKarczmarekNogueiraVanRaamsdonk, Wall2, HubenyRangamani, BoussoLeichenauerRosenhaus} for discussion of this question. 

One should bear in mind that extremal surfaces are not the only probe used in AdS/CFT.  If there are other bulk probes besides extremal surfaces, it is possible that a firewall (if it exists at all) might be located somewhere behind the outermost barrier.  Conversely, if there are some extremal surfaces which do not actually correspond to dual observables (for example, because they are not the \emph{minimum} area extremal surface), then it could conceivably be that the firewall is actually outside of the barrier.  It may also be that there are regions of space where \emph{some} kinds of extremal surfaces can probe, but not enough to fully reconstruct the geometry.  It is unclear what the status of these regions would be. 

In the case of a Vaidya spacetime where a black hole forms from the collapse of a shell, there is no extremal surface barrier.  In particular, there are spacelike geodesics which pass through any point of the black hole interior.  The AMPS argument for firewalls \cite{AMPS} applies to sufficiently old black holes that form from collapse, yet the absence of a barrier might seem to suggest that there can be no firewall.  However, the spacelike geodesics in question are anchored at very large time separation on the boundary.  In fact, \cite{LoukoMarolfRoss} argued that the WKB stationary phase approximation is not dominated by spacelike geodesics passing through the interior of the black hole, but rather by complex geodesics. It is unclear how to use the barrier results proven here in the case where the extremal surfaces are complex, as was found in \cite{FidkowskiHubenyKlebanShenker} for two-point correlators in the case of an AdS-Schwarzschild bulk.   


However, even for the Vaidya spacetime, we still find that there is a partial barrier, for spacelike surfaces anchored sufficiently far to the future of the collapsing shell.  This is because the spacetime is just Schwarzschild at late times.  Thus the extremal surface barrier might still give clues about the location of the firewall, although it does not necessarily tell us when the firewall would first appear.

In general, our knowledge of the AdS/CFT dictionary comes from a bootstrapping procedure where (a) proposed new ingredients to the dictionary can be checked using our knowledge of bulk physics, while (b) the behavior of the bulk can also be predicted using the dictionary.  But if at step (b), our current understanding of the dictionary predicts a firewall---resulting in a breakdown of the bulk equations of motion used at step (a)---then it is unclear how to proceed.  Should we modify the bulk equations of motion or the dictionary?  Perhaps it is the local bulk equations of motion which should be kept sacrosanct, and other things we think we know about AdS/CFT should be modified in order to preserve them.  Ultimately, the question is what choice of dictionary leads to the most consistent, complete, and interesting form of the correspondence.

There are several remaining questions worth noting.  First, we have yet to find a complete description of the most general barrier, in particular for the outermost barrier surface. Such a description would facilitate an understanding of precisely how generic extremal surface barriers are.  We expect that, assuming the null curvature condition, it should be possible to show that compact outermost barriers can only occur in spacetimes with singularities.

We note that (aside from Theorem 2.5) the main results of this paper are purely geometrical in nature---they apply to any Lorentzian spacetime without needing to use of any energy condition restricting the sign of the curvature.  (Although given the existence of trapped surfaces as shown in Theorem 4.2, the null curvature condition would say that there must be singularities.)  This suggests that our results are still valid even in semiclassical spacetimes (e.g. evaporating black holes) where local energy conditions are violated.  However, the significance of these geometrical results for AdS/CFT is not the same, since in the semiclassical (finite $N$) regime, there are quantum corrections to the dictionary relating extremal surfaces to boundary physics.

For example, the entanglement entropy of a CFT region is no longer given just by the area of an extremal surface.  Instead it receives corrections due to the bulk entanglement entropy \cite{FaulknerMaldacena}.  Thus we should really be interested in surfaces which extremize the area \emph{plus} the bulk entanglement entropy.  To find barriers to these surfaces, it is natural to look for null surfaces foliated by quantum trapped surfaces, where the sum of the area and the bulk entropy is decreasing.  One can then prove quantum generalizations of the types of classical theorems proven here \cite{Wall}.
\\



\vskip 2cm
\centerline{\bf Acknowledgements}
\vskip 1cm
\noindent We thank Dalit Engelhardt, Sebastian Fischetti, Ahmed Almheiri, Gary Horowitz, Don Marolf, William Kelly, and Mudassir Moosa for helpful discussions. This work is supported in part by the National Science Foundation under Grant No. PHY12-05500. NE is also supported by the National Science Foundation Graduate Research Fellowship Program under Grant No. DGE-1144085. AW is also supported by the Simons Foundation.

\end{document}